\documentstyle[12pt,twoside,fleqn,epsf,espcrc1]{article}

\newcommand{\AmS}{{\protect\the\textfont2
  A\kern-.1667em\lower.5ex\hbox{M}\kern-.125emS}}

\title{From Notes to Chords in QCD}

\author{Frank Wilczek\address{Institute for Advanced Study, 
        School of Natural Science, \\ 
        Olden Lane, Princeton, NJ 08540}
        \thanks{Currently on leave at Leiden University.
        Instituut-Lorentz, Niels Bohrweg 2, 2333 CA Leiden, The
        Netherlands.  Supported in part by DOE grant
        DE-FG02-90ER40542}}  
        
\begin{document}
\maketitle

\begin{abstract}
After a very brief overview recollecting the `classic'
parts of QCD, that is its application to describe hard processes and
static properties of hadrons, I survey recent work -- some very recent
-- on QCD at non-zero temperature and density.  At finite temperature
and zero density there is a compelling theoretical framework allowing
us to predict highly specific, non-trivial dependence of the phase
structure on the number of flavors and colors.  Several aspects have
been rigorously, and successfully, tested against massive numerical
realizations of the microscopic theory.  The theoretical description
of high density is nowhere near as mature, but some intriguing
possibilities have been put forward.  The color/flavor locked state
recently proposed for three flavors has many remarkable features
connected to its basic symmetry structure, notably including chiral
symmetry re-breaking and the existence (unlike for two flavors) of a
gauge invariant order parameter.  I survey potential applications to
heavy ion collisions, astrophysics, and cosmology.  A noteworthy
possibility is that stellar explosions are powered by release of QCD
latent heat.
\end{abstract} 



\bigskip
\bigskip

One main goal of physics is to determine the fundamental laws, where
``fundamental'' is taken in the strict reductionist sense, that is,
laws incapable of being derived from other, more universal principles.
I believe it is profoundly wrong, however, to portray this as the only
goal, or even necessarily as the most important goal.  It is as if
after playing each note of the piano once each, you were to say ``and
all the rest is just combinations''.  True enough -- but somehow
misleading.  In QCD we first played the basic notes twenty-five years
ago, and certainly by fifteen years ago it was clear to most
reasonable people that there were no more notes to be found.  But it
would be quite foolish to see this as the end of the subject.  Indeed,
several interesting, attractive chords have been discovered already,
and large parts of the keyboard have barely been touched.



\section{Pure Tones at High Frequency: Jets, Running of Couplings, Proton
Gluonization }


The fundamental structures of QCD, both as regards its degrees of
freedom and their dynamics, are laid bare in the phenomena of jets.
Asymptotic freedom says that radiation events involving large changes
in energy and momentum are rare.  Thus when highly energetic quarks
and antiquarks are produced, say following $e^+e^-$ annihilation at
high energy, including the important special case of $Z$ production,
they imprint their pattern of energy and momentum flow.  Accordingly
one expects that soft radiation events can rearrange the particle
types, and of course ensure that the quarks, antiquarks, and gluons
materialize as hadrons, but that the underlying energy-momentum
structure remains visible, now as jets of hadrons instead of
individual quarks.  Hard radiation events, although relatively rare,
do happen, and in that case one has three jets, or, still more rarely,
four or more.  All aspects of these events -- the relative
cross-sections for different numbers of jets, how this varies with
energies, and the angle and energy distributions -- can be calculated
directly from the microscopic theory.  In this way, the precise
structure of the fundamental quark-gluon and gluon-gluon couplings has
been rigorously, and successfully, tested in detail.

Jets provide perhaps the most striking and direct demonstrations of
the Yang-Mills structure of QCD and of asymptotic freedom, but these
were by no means the first test of the theory, nor are they the most
accurate.  A very wide variety of experiments at various energy scales
has been used to test the theory \cite{schmel}.  In particular, the predicted
running of the coupling has been convincingly observed (Figure 1).

\begin{figure}[htb]
\begin{center}
\hspace*{0pt}
\epsfysize=80mm
\epsffile{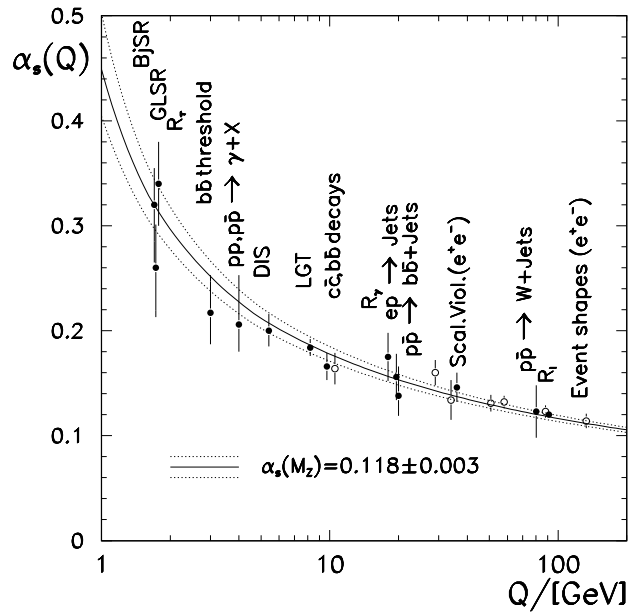}
\end{center}
\caption{Comparison of theory and experiment in QCD,
illustrating the running of couplings.  Several of the points on this
curve represent hundreds of independent measurements, any one of which
might have falsified the theory. Figure from M. Schmelling, hep-ex/9701002.}
\label{fig:asdrun.eps}
\end{figure}

Before leaving this subject I would like to especially mention one of
the most dramatic, and to me most gratifying, experimental
verifications of QCD.  This is the rise in the nucleon structure
functions at small x, now clearly observed at HERA \cite{h1collab}.
These large -- factor 2 or more -- violations of naive scaling were
among the very first consequences derived from asymptotic freedom
\cite{derujula}.  Regrettably, some of the early analysis of this data
got enmeshed with murky speculations about hard pomerons and presented
as a duel between inappropriate acronyms.  Further work has made it
clear that the original ideas and calculations were right on the
money.


 
\section{Pure Tones at Low Frequency}


Since we will be considering the effects of having different numbers
of flavors, there is one additional basic concept worth bearing in
mind.  It is quite far removed from realistic phenomenology, but might
provide us with a nice theoretical laboratory for special questions,
as I will discuss further below.  For large enough numbers of quarks
the coupling never gets strong, and one can calculate the behavior
both at small and at large distances very simply.  In addition to the
usual ultraviolet stable zero of the $\beta$ function at zero
coupling, there is an infrared stable zero at non-vanishing coupling.
When the number of massless quarks is close to 16 this second zero
occurs at a small value of the coupling, and one can use perturbation
theory at all scales \cite{dgross}.  The theory becomes scale
invariant and free at short distances, and scale invariant again with
non-trivial exponents at long distances.  There is no chiral symmetry
breaking and no mass gap.  The same qualitative pattern might persist
to smaller numbers of quarks, as in supersymmetric versions of QCD
\cite{seiberg}, but the infrared fixed point will move to strong
coupling, making accurate quantitative calculation much more
difficult.



\section{Hadrons} 



For more realistic quark spectra, and in the real world, the coupling
becomes large, and a mass gap appears, at some characteristic scale
usually denoted $\Lambda_{\rm QCD}$.  Of course, there can be a few
Nambu-Goldstone modes appearing below this scale.

As you can see from Figure 1, the effective coupling has a focusing
property: for a wide range of $\Lambda_{\rm QCD}$, one obtains very
nearly the same value of the effective coupling at $M_Z$.  This
realizes, in the context of QCD, Pauli's dream of calculating the fine
structure constant.  Concretely, it means that one has formulated a
wealth of quantitative predictions for physical phenomena, in which no
material parameters whatsoever appear.

A clearer view of this remarkable aspect of QCD appears if we consider
the idealized limiting theory containing just two massless quarks (the
idealized $u$ and $d$).  On the one hand, this idealization provides a
reasonably accurate representation of the ``everyday world'' of
nuclear physics and of the physics of non-strange resonances, probably
at the 10\% level, apart from those few phenomena that depend
sensitively on the pion masses.  On the other hand, it is a theory
whose equations, when expressed in units with $\hbar = c = 1$, contain
no free parameters at all, apart from the pure numbers 3 (colors) and
2 (flavors).  Classically there is a gauge coupling $g$ that appears to be 
an
arbitrary number, but the running of the coupling implies that the
physical coupling changes with mass or distance scale, and we can work
with any value we like, by choosing the scale appropriately.  This is
the phenomenon of dimensional transmutation.  Thus everything is fixed
and calculable in terms of the units $\hbar$, $c$, and the whole
numbers 3 and 2, except for the overall scale of mass, which in any
case has no intrinsic meaning for QCD as such.  QCD is a theory
Pythagoras (``All things are numbers'') would have loved.

I feel it is vitally important for a scientific theory 
to have algorithms
in hand to deliver some concrete returns on its abstract promises.
Without such algorithms, the promises are hollow.  Fortunately, the
asymptotic freedom of QCD allows us to discretize this theory in a
controlled manner.  By putting the fields on a lattice \cite{wilson}
one is throwing out the high-frequency modes, but these modes are
weakly coupled, so one can assess their residual effects accurately
and approach the limit of vanishing lattice spacing with confidence.
    
After years of hard work, much of it pushing the frontiers of
computing technology, heroic practitioners of numerical QCD have not
only made it clear that the microscopic theory generates the
qualitative phenomenology of confinement \cite{creutz} and chiral
symmetry breaking, but also produced a convincing approximation to the
low-lying spectrum, at the 10\% level \cite{butler}.


Much work remains to be done in lattice gauge theory, to improve the
accuracy of results for static quantities and to calculate matrix
elements of interest for analyzing weak hadronic processes.  In
addition, there are challenges of a more conceptual sort: Can one
incorporate manifest chiral symmetry in a usable non-perturbative
algorithm?  Can one isolate appropriate degrees of freedom to obtain a
semi-microscopic, more easily tractable version of QCD?  There are
some promising ideas for both classes of problems -- notably domain
wall fermions \cite{kaplan}, and the instanton or monopole-instanton
liquid \cite{shuryak} -- that need to be pushed harder.  Finally and
perhaps most important, there are questions that touch fundamental
phenomenological issues: What is the mass of the H particle?  Are
there stable or metastable strangelets \cite{bodmer}?  Is it possible
that the mass of the $u$ quark vanishes, solving the strong P, T
problem?  In principle QCD is adequate to address these issues, but
they remain open due to our weakness in calculation.


\section{Finite Temperature}


As a heuristic principle, asymptotic freedom leads us to expect that
at high temperature one should approach the behavior of the free
theory.  In particular, confinement should no longer apply and chiral
symmetry should no longer be broken.  These qualitative changes
entail, in favorable cases, strict phase transitions.


It is instructive first to consider the pure glue theory, both for
$SU(2)$ and $SU(3)$.  In the pure glue theory there is a strict
criterion for confinement, namely the vanishing of the Polyakov loop.
This loop represents the effect of inserting a source of non-zero
diality (for $SU(2)$) or triality (for $SU(3)$).  Unbroken symmetry
under the global transformations associated with the center of the
gauge group, $Z_2$ or $Z_3$ respectively, requires the expectation
value of the Polyakov loop to vanish.  This can be interpreted,
alternatively, as a sign that the flux associated with the source
cannot be screened.  Its influence extends to infinity, and it
produces a disturbance costing a finite energy per unit volume over an
infinite volume, hence an infinitely massive state.  At high
temperature there will be screening, the Polyakov loop will not
vanish, and the global center symmetry will be violated.  Since there
is a change of symmetry, there must be a sharp phase transition
somewhere between zero and infinite temperature.

The next standard question to ask is the order of the phase
transition.  A powerful approach to this problem is to consider
whether it is possible for the transition to be second order.  The
singularities at second order transitions are supposed to be governed
by scale invariant theories at the critical point, and to have a
universal character.  Thus one can analyze much simpler models than
QCD, that have the appropriate symmetries, while obtaining results
that apply quite rigorously to QCD.  The long-wavelength modes, that
alone are important for the singular critical behavior, should not
depend on the microscopic details.

For $SU(2)$, with its center $Z_2$, an appropriate model is the three
dimensional Ising model \cite{svetitsky}.  It has a second order
transition, with critical exponents that have been accurately
calculated.  The critical behavior of pure glue $SU(2)$ has been
calculated quite accurately.  The critical exponents agree with those
of the 3 dimensional Ising model, within 1-2\%.  This is a remarkable
example of universality \cite{engels}.

For $SU(3)$, with its center $Z_3$, an appropriate model is the
3-state Potts model.  The cubic coupling allowed in this universality
class inevitably leads to a first-order transition.  Thus pure glue
$SU(3)$ is predicted to \cite{svetitsky}, and does \cite{engels}, have
a first-order deconfining transition.
 

For the more realistic case of gauge theory with dynamical quarks,
there is no strict symmetry associated with confinement; at any
non-zero temperature, the quarks will screen triality, so that the
Polyakov loop will never strictly vanish and the global central gauge
symmetry is always broken.  However if we have more than one massless
quark species there are non-trivial chiral symmetries, and we can
apply the same style of analysis.

For two flavors the chiral symmetry is $SU(2)_L\times SU(2)_R$, which
is broken at zero temperature to the diagonal $SU(2)_{\rm L+R}$.  In
isomorphic but more suggestive notation the breaking is $SO(4)
\rightarrow SO(3)$, and we recognize the universality class of a
four-component magnet.  This universality class has been closely
studied both analytically and numerically.  It has a second order
transition, with accurately determined exponents and critical equation
of state.  Using this information, one can test the hypothesis that
two massless flavor QCD is in the universality class \cite{pisarski}.
In very beautiful work, Iwasaki, Kanaya, Kaya, and Yoshie
\cite{iwasaki} have, to my mind quite convincingly, demonstrated that
it is.  They were able, for example, clearly to distinguish a fit to
their data using the $SO(4)$ exponents ($\chi^2/df = .72$) from one
using mean field exponents ($\chi^2/df \geq 3.3$).

For three flavors the chiral symmetry is $SU(3)_L \times SU(3)_R$,
which is broken at zero temperature to the diagonal $SU(3)_{\rm L+R}$.
The simplest model in this universality requires a 3x3 matrix of
scalar fields, and has two independent quartic couplings.  The
infrared fixed point present in the mean field theory is unstable when
fluctuations are considered.  Thus there is no candidate model for a
second order transition with the appropriate symmetry, and one
predicts a fluctuation-driven first order transition \cite{pisarski}.
Numerical experiments with three flavors of massless quarks confirm
this expectation \cite{kanaya}.

For more flavors one also expects a first order restoration of chiral
symmetry, if it was broken at zero temperature, for the same reason.
This is also consistent with the numerical experiments.


Since two and three massless flavors give such different pictures, and
it is not clear {\it a priori\/} which is a better approximation to
reality, it becomes important to consider how the two pictures might
morph from one to the other.  Concretely, one can imagine starting
with vanishing strange quark mass, and cranking it up.  At the
beginning of this process we have a first order transition, and at the
end a second order transition.  The simplest possibility is that the
transitions are continuously connected.  The switch happens at a
tricritical point, which has universal properties that can be
rigorously characterized \cite{wilczek}.  Available numerical data, although
sparse, is consistent with this picture.  Another question is the
nature of the transition, that is first or second order, for the
physical value of the strange quark mass.  Different groups studying
this question have reached opposite conclusions \cite{fbrown,kanaya}, which may
indicate that reality is close to tricritical!
 
Of course in reality the $u$ and $d$ quarks are not strictly massless.
This will round the second order transition into a crossover, and
change the tricritical point into an Ising-like transition.  However
insofar as the light quark masses are small it will be fruitful to
regard them as perturbations within the more symmetric framework,
joining onto a small region very near the critical point where they
dominate.


There are many other important issues that can be addressed in finite
temperature QCD, besides the universal features of the phase
transitions.  Let me just mention the simplest one, the transition
temperature.  First of all, one finds that the ratio of string tension
to transition temperature is markedly different for the pure glue and
dynamical quark theories, presumably reflecting the fundamentally
different character of the light degrees of freedom, and of the
transition, in these two cases, despite the fact that the gluons
overwhelmingly dominate the short-distance dynamics.  The actual value
of the critical temperature is of course very important for
phenomenology. For the dynamical quark theory, it appears to lie
within the range $T_c \sim$ 150-200 Mev \cite{tblum}.  It is striking that
this value is so small.  Just below the transition one has essentially
only pions -- three degrees of freedom; while above, if one imagines a
free phase, there are fifty-two (eight gluons with two helicities
each, plus three flavors of quarks each with three colors and two
helicities, and their antiparticles)!
   
\section{Finite Density}

We can form reasonable expectations for the behavior of QCD at very
high densities.  At high density -- assuming weak coupling -- the
quarks will form a large Fermi surface.  We want to see if this trial
state, which is embedded in a nearby continuum, is stable.  We are
doing nearly degenerate perturbation theory, involving small numbers
of particle-hole excitations near the Fermi surface.  Since other
possibilities are blocked by the exclusion principle, the leading
interactions are elastic scatterings among these particles and holes.
Because the momenta are large, such scatterings, over the bulk of
possible angles, involve large momentum transfers.  By asymptotic
freedom, they are characterized by weak coupling, so our starting
point is self consistent.  Singular infrared behavior is eliminated by
screening, so I suspect that the neglected ``small angle''
contributions are truly negligible asymptotically, and the rough
argument sketched here can be made fully rigorous, but this has not
been attempted.  In any case for physical purposes we are not going to
be satisfied with asymptotics, and we shall have to dirty our hands.

So what is it reasonable to expect?  Ordinary chiral symmetry
breaking, as seen at zero density, is a quark-antiquark pairing.  It
will become ever more difficult to maintain at high density, because
the `quark' parts of the pairs for low momentum are definitely
occupied in the Fermi sea, and their number cannot vary, so there is
no possibility for correlated pairs.  And at high momentum, the
quark-antiquark pairs have large energy.  On the other hand, there are
attractive quark-quark channels.  Indeed, a quark pair in a color
antitriplet has reduced its flux, and should be lower in energy than
the two quarks separately.  The energy denominators in this channel
vanish near the Fermi surface.  Since a BCS-type instability can be
triggered by an arbitrarily weak interaction at the Fermi surface, we
expect pairing condensations of some sort in the quark-quark channel
\cite{bailin}.

A fundamental question for QCD is the nature of the transition that
restores chiral symmetry.  It is very attractive, on general physical
grounds, to believe that it is first order.  To see one reason why,
imagine setting up a medium of uniform small density with chiral
symmetry broken.  If the chiral symmetry restoration is first order,
this medium will be mechanically unstable.  It will break up into
regions of non-zero density wherein chiral symmetry is restored, and
regions of zero density where chiral symmetry is broken.  This seems
very reasonable, and indeed this picture could be used to motivate,
and provide a rigorous approach to, something like the MIT bag model.
On the other hand if the uniform medium were stable -- what could it
be?  I don't know of any candidates in the physical world.  Numerical
work based on effective instanton interactions, in agreement with some
but not all other approaches, supports the idea of a first-order
transition \cite{malford}.

Now let us consider what's on the far side of the transition,
asymptotically at very large density, keeping in mind that there could
be one or more intervening transitions (e.g. K condensation) near
nuclear density. Once again, it is important (and instructive) to
consider different numbers of flavors.

For two  flavors,  triplet pairing of the form
\begin{equation}
\langle q^\alpha_a (p) C\gamma_5 q^\beta_b (-p) \rangle ~=~
\kappa (p^2)  \epsilon^{\alpha\beta 3} \epsilon_{ab}
\end{equation} 
is possible \cite{malford} \cite{rapp}.  
It puts the affected quarks (colors
1 and 2) in a color antitriplet, space-time scalar s-wave condensate.
We might expect it to be especially favorable because it maintains so
much symmetry, so that the interaction of a given pair can receive
coherent contributions from other pairs involving different particle
combinations all over the various Fermi surfaces.  Indeed, it breaks
the original color $SU(3)$ down to $SU(2)$ while leaving chiral
$SU(2)_L\times SU(2)_R$ intact.  It might seem that the $U(1)$ of
baryon number is violated, but actually there is a combination of the
original baryon number and color hypercharge which leaves the
condensate invariant, so a modified global $U(1)$ persists.  Estimates
of the energetics using effective instanton interactions make it
plausible that substantial gaps, of order 100 Mev or more, can be
generated at moderately high chemical potentials, corresponding to
densities a few times nuclear.  The corresponding critical temperatures
range from several tens to perhaps 100 Mev \cite{berges}. 

This pairing leaves quarks of the third color untouched, and one can
ask whether the residual ungapped Fermi surface is stable.  In our
crude effective instanton model \cite{malford} we found that there
could be an additional condensation of the form
\begin{equation}
\langle q^3_a(p) C\sigma_{0i} q^3_b(-p) \rangle = \eta (p^2) {\hat
p}_i \delta_{ab}~.
\end{equation}
This violates rotation
symmetry.  Condensation in this channel is not very
robust and gave us gaps of order several keV at best.

For three flavors, there is what I think is a much more beautiful and
compelling possibility for condensation \cite{raja}, of the form
\begin{equation}
\langle q^\alpha_a(p) C\gamma_5 q^\beta_b(-p) \rangle ~=~
\kappa_1 (p^2) \delta^\alpha_a \delta^\beta_b +
\kappa_2 (p^2) \delta^\alpha_b \delta^\beta_a~.
\end{equation} 
This can be written so it appears more 
obviously as a generalization of the
2 flavor condensation, by noting that if $\kappa_1 = - \kappa_2$ then
the color-flavor structure is proportional to
$\epsilon^{\alpha\beta I} \epsilon_{abI}$, summed over $I$.  It is
reminiscent of the B phase of liquid helium 3, where now color and
flavor rather that nuclear and orbital spin become locked together.
We have found, in an effective model abstracted from one-gluon
exchange, that this condensation leads to very substantial gaps, 
easily of order 100 Mev or more.

A state with this condensation has many remarkable features:

The symmetry
$SU(3)_{\rm color}\times SU(3)_L \times SU(3)_R \times U(1)_{\rm L+R}$
of color times chiral $SU(3)$ times baryon number breaks down to the
diagonal $SU(3)$.

In particular, chiral symmetry is broken by a mechanism that is
qualitatively new.  That is, the left-handed quarks lock their flavor
quantum numbers to color and so, essentially independently, do the
right-handed quarks.  But because the kinetic terms 
in the Hamiltonian are only invariant under vectorial color
transformations, left and right chiral symmetries become
locked to each other, indirectly.

The $U(1)$ of baryon number is genuinely broken.  This leads to
superfluidity, as in liquid helium.  Since baryon number is -- for
present purposes -- an exact symmetry of the Hamiltonian, its
spontaneous violation is associated with a true Nambu-Goldstone mode,
with a linear dispersion relation at small momenta (but velocity equal
to the speed of zero sound, close to $c/\sqrt 3$).

There are gaps in all single-particle channels, {\it i.e}. everywhere
around all the Fermi surfaces.

There are approximate Nambu-Goldstone modes associated with the
spontaneous chiral symmetry breaking.  When quark masses are included,
these modes acquire small gaps too.  Then all the charged modes, both
for weak and electromagnetic interactions --
but not the neutral one mentioned above -- have gaps.

All the gluons acquire a mass, essentially by the Higgs mechanism.

Although ordinary electromagnetic gauge invariance is broken, there is
a combination of this transformation and color hypercharge that leaves
the condensate invariant.  Under this transformation, all the
elementary excitations -- quarks, gluons, and
collective modes -- carry integer charge (yes, in multiples of the
electron charge)!  Furthermore since there is a gap to all charged
excitations, this ultra-dense material is transparent, like diamond.

It is a very interesting enterprise, to derive the Landau-Ginzburg
effective description of the low energy excitations around this state.
By the way it also supports Skyrme-type textures, which carry baryon
number (measured by the effective coupling
of the Nambu-Goldstone mode!) equal to 2.

\bigskip

For very large numbers of flavors, I am not sure what happens --
perhaps multiple copies of the color-flavor locked state for groups of
three quarks, with some fudge for any residuals.  In any case when the
number gets close to 16 it is a truly weak
coupling problem at any finite density, and one should be able to
make progress analytically.

In this discussion I have been rather cavalier in dealing with
gauge-dependent quantities.  The same difficulty arose, and was faced
down, in BCS theory \cite{anderson}.  In the pairing wave function, taken at
face value, charge is not conserved, whereas of course in reality it
had better be.  This formal problem can be ameliorated by projecting
on a state of definite number, by doing a Fourier transform on the
conjugate phase variable $\theta$.  This operation will have little
practical effect, in a large system, because states with
significantly different values of $\theta$ have very little overlap,
nor do local operators constructed from reasonable numbers of field
operators connect them.  The important physical phenomenon captured in
the pairing wave function is the correlation among however many pairs
do exist, and this is not affected by global constraints on the
(large) number of pairs.  The same remarks apply, {\it mutatis
mutandis}, for color superconductivity.

As a consequence of the projection on color singlets, however, naive
attempts to characterize ordered states using formally defined order
parameters that are not gauge invariant will fail.  The primary stage
of our two flavor diquark pairing apparently does not support any gauge
invariant order parameter. In this respect it resembles the
electroweak sector of the Standard Model.  As in that case, the
logical possibility of a crossover rather than a sharp transition to
nearby states with the same symmetry but qualitatively different physical 
behavior thereby arises.

On the other hand, there are gauge invariant order parameters both for
the complete (two-stage) two flavor condensation and, importantly, for
the three flavor color-flavor locking 
condensation. In the latter case, an appropriate 
gauge-invariant order parameter is the expectation value of a six-quark 
operator of the type $(qqq)^2$, which captures the spontaneous $U(1)$ 
violation.  States supporting 
these condensations must therefore be separated from the 
surrounding states, having different symmetry, by sharp phase transitions.

With so many interesting possibilities being suggested for various
versions of QCD at finite density, it would be very desirable to have
the kind of definitive checks that only numerical realization of the
theories can provide.  Unfortunately, there are very great
difficulties with numerical work on QCD at finite density, which have
for many years now prevented the subject from getting off the ground.
The fundamental problem is that the Fermion determinants appearing in
the functional integral measure for the grand canonical ensemble are
not positive definite configuration by configuration.  Thus importance
sampling is stymied.  One can use the measure associated with zero
chemical potential, and evaluate the expectation value including 
chemical potentials, but this has been found to converge very
(exponentially)  slowly due to cancellations of many unwanted zero
density contributions, which are much larger in absolute size than the
residual terms of interest.  The deeply nasty character of this 
problem has only been appreciated recently \cite{barbour}.

One can avoid the problem in various ways.  One class of ideas is to
work with models that don't suffer from it, but still do share some
features with QCD.  For example with two colors the problem doesn't
arise \cite{knowledge}.  Using two colors is far from innocuous in the
context of finite density QCD, because it makes the baryons into
bosons, but one might hope to learn something.  Or one can have a
chemical potential which is positive for some quarks and negative for
an equal number of others \cite{kapustin}.  This will favor production
of mesons with appropriate quantum numbers, as well as baryons (and
antibaryons), so it too is far from innocuous.  Or one can look at
models which have qualitative features in common with QCD but do not
suffer from the sign problem -- the 2+1 dimensional Gross-Neveu model
is a good candidate \cite{shands}.  Or one can use an imaginary
chemical potential \cite{kapustin}.  This I think is a particularly
interesting possibility, and I'd just like to say a few words about
when it is likely to be useful, since the considerations apply more
generally.  An imaginary chemical potential does not systematically
bias the ensemble to large density, so that to pick out the effect of
states of non-zero baryon number density one must rely on fluctuations
(which, when they occur, are appropriately weighted by the action).
These fluctuations will occur most readily when the temperature is
high, and the gap in the baryon number channel is small. And even then
one can only realistically hope, on the small lattices likely to be
practical, to fluctuate to a few baryons.  So a reasonable procedure
would seem to be to start with a high temperature and work down,
looking for qualitative changes as a function of temperature.  In this
way, one could realistically hope to study how properties of baryons
are affected by the presence of a thermal meson medium, and in
particular, the transition from a loose association of almost massless
quarks to something resembling a normal baryon.

Another possibility is to work with large numbers of quark species,
close to 16.  Then there is no mass gap to baryons, so fluctuations
are cheap, and also the contribution of interest, due to the quarks,
is not swamped by gluons.  So the cancellations are probably not so
bad even at real chemical potential, and the imaginary chemical
potential approach should also work better (since the fluctuations
will be cheap).

Still, these tricks are not going to help us directly on realistic
problems.  For those, new ideas of a more profound character seem to
be required.  Perhaps the condensed matter theorists, who have been
wrestling with related sign problems increasingly successfully
recently, have something to teach us here.


\section{Phenomenology}


Finally I'd like to discuss some potential phenomenology associated
with QCD in extreme conditions.

There is a very large literature concerned with various ways of
establishing, phenomenologically, that a shift from hadronic to
quark-gluon degrees of freedom as the appropriate description of
strong interaction physics takes place under conditions of high
effective temperature \cite{hmeyer}.  One good idea in this field
\cite{matsui}, which has already received significant experimental
support \cite{na50}, is that the liberation of gluons from hadrons
hardens their distribution function, permitting them to dissociate
J/$\psi$ particles more readily.  Closely related to this is the
suggested phenomenon of hard jet suppression \cite{bjorken}.  Also
impressive, though less clear-cut, are suggestions of strangeness and
entropy enhancement in high energy nuclear collisions, already visible
in existing data.  I think there is little doubt that by correlating
various signals of these and other sorts one will be able to construct
a convincing, richly detailed case that at high temperatures the
description of strongly interacting matter as semi-free quarks and
gluons is appropriate and fruitful.

Within the domain of heavy ion physics, I'd like to advertise three
specific phenomena that could provide signatures and guidance for
elucidating more subtle, collective aspects of the behavior.

{\it DCC and anomaly kick}: The idea that as the chiral condensate
first melts and then relaxes it puts excess power into long-wavelength
modes has been developed over the past several years, and there are
excellent reviews available \cite{krishna}.  Recently a promising new idea was
introduced by Asakawa, Minakata, and Muller \cite{asakawa}, which apparently
makes some associated phenomenology much more robust and accessible
than it previously appeared.  This is the idea of the anomaly kick.
The Adler-Bell-Jackiw anomaly gives a coupling of the $\pi^o$ to the
product of electric and magnetic fields, and in a heavy ion collision
large coherent electric and magnetic fields are present, simply due to
the charge and motion of the nuclei.  By including this term in the
effective field theory, one seeds the long-wavelength modes in the
$\pi^o$ direction.  If these are then amplified in the predicted
manner, a substantial, systematic enhancement of low-momentum neutral
pions is expected. It can, of course, and should be measured relative
to the charged pion background.

{\it Tricritical point}: In recent, beautiful work Stephanov,
Rajagopal, and Shuryak \cite{stephanov} pointed out that a
continuation of tricritical point we discussed above -- or, in view of
non-vanishing light quark masses, simply critical point -- could be
experimentally accessible.  We accessed it notionally by dialing the
strange quark mass; these authors point out that one might hope to
access it experimentally, by using instead baryon density (in addition
to temperature) as a control parameter.  We are speaking here of a
true critical point, involving long-range correlations.  SRS suggested
a number of specific, striking observable phenomena that could allow a
convincing identification of the predicted critical point.

{\it Flavor flow}:  Although it has not yet been properly quantified,
several of us have for some time been discussing the possibility
that tendencies for formation of condensates with non-trivial flavor
structure could induce observable flavor flow.  Probably the most
hopeful, and certainly the most fundamentally interesting, case
arises for the color-flavor locked state.  In a heavy ion collision
one starts with essentially no strange (or antistrange) quarks.  If
condensation of the color-flavor 
locking type is energetically favorable, at high density --
preferably, at low temperature -- one could find it favorable to
rearrange strange-antistrange pairs, as they are produced, so the the
strange quarks are retained in the condensate, while the antistrange
are allowed to drift away.  In the extreme case, it could even become
favorable to produce pairs spontaneously classically, if the rest
mass of the strange pair is made up by condensation energy.
Qualitatively, the signature would
be a concentration of low-momentum strange quarks -- evolving into
strange baryons -- in the central {\it collision\/} region, and a
compensating broad enhancement of antistrangeness flowing out of it.

Turning to astrophysics, I'd like to mention two possibilities, one
relatively conservative, the other potentially of dramatic consequence
for the theory of stellar explosions.  The (relatively) conservative
possibility is that quark matter deep inside neutron stars is in the
color-flavor locked state.  The major consequence of this is that
there is a gap to all charged excitations, so that it is difficult to
lose heat by normal electromagnetic or weak processes, while there is
no gap to the superfluid modes associated with spontaneous baryon
number violation, so that there is a repository for heat.  The
qualitative effect of this would be to keep neutron stars warm for a
longer time than was otherwise possible with quark matter.

A more dramatic possibility is opened up by possibility, arising
perhaps because of color superconductivity and almost certainly from
color-flavor locking, that QCD makes a transition to a {\it highly
ordered\/} state at high density, and one having radically different
properties with respect to the weak interaction.  
This state could be reached
directly from compressed nuclear matter, or (perhaps more likely)
through an intermediate phase with chiral symmetry restoration.
Transition to the ordered state will be accompanied by release
of latent heat.  The available energy, being naturally measured in
terms of the basic interactions responsible for the bulk of nucleon
masses, could be very substantial on the scale of nuclear
astrophysics.  It could provide the key to the whole problem of
getting collapsing stars to explode -- a problem that has had a long,
checkered history, and continues to lack a clear, robust solution.
This inspiring possibility deserves much further attention.

Let me emphasize that these possibilities do not necessarily conflict
with one another, nor need they be connected with the more radical
\cite{bodmer}, possibly problematic \cite{caldwell}, idea that an
exotic form of strongly interacting matter is the ground state at zero
pressure.

Turning from astrophysics to cosmology, the main message form QCD
studies appears to be the satisfactory though perhaps disappointing
one, that no dramatic relics of the QCD phase transition seem likely
to survive processing through the big bang.  I say this is
satisfactory, because there is no hint of any such relic,
e.g. inhomogeneous nucleosynthesis or exotic baryonic dark matter.  It
might have been otherwise, if the phase transition at high density and
nearly zero chemical potential were strongly first order.  Schmid,
Schwarz, and Widerin \cite{cschmid} have pointed out that the modification of
the equation of state during a QCD phase transition or crossover,
specifically the pressure deficit, could enhance the growth of
fluctuations in kinetically decoupled dark matter (e.g., axions) on
small scales.  If this is a significant quantitative effect, it could
have a drastic, negative effect on prospects for axion detection in
ongoing and planned experiments.  To decide this question for sure, we
need better control of the equation of state through the phase
transition or crossover.


\begin{thebibliography}{9}
\bibitem{knowledge} This is common knowledge among experts in the
field, but I have not been able to find a proper reference.
\end{thebibliography}
\end{document}